\documentclass[journal]{IEEEtran}

\usepackage{cite}

\usepackage{amsmath,amssymb,amsfonts}

\usepackage{graphicx} 
\usepackage{xcolor}

\begin{document}

\title{Fifteen Years of Progress at Zero Velocity: \\A Review}

\author{Johan Wahlstr\"{o}m and Isaac Skog 
	\thanks{J. Wahlstr\"{o}m is with the Dept. of Computer Science, University of Exeter, Exeter, EX4 4QF, UK (e-mail: j.wahlstrom@exeter.ac.uk).}
	\thanks{Isaac Skog is with the Dept. of Electrical Engineering, Link\"oping University, 581 83 Link\"oping, Sweden (e-mail: isaac.skog@liu.se).}
}

\markboth{}%
{Shell \MakeLowercase{\textit{et al.}}: Bare Demo of IEEEtran.cls for IEEE Journals}

\maketitle

\begin{abstract}
Fifteen years have passed since the publication of Foxlin's seminal paper ``Pedestrian tracking with shoe-mounted inertial sensors''. In addition to popularizing the zero-velocity update, Foxlin also hinted that the optimal parameter tuning of the zero-velocity detector is dependent on, for example, the user's gait speed. As demonstrated by the recent influx of related studies, the question of how to properly design a robust zero-velocity detector is still an open research question. In this review, we first recount the history of foot-mounted inertial navigation and characterize the main sources of error, thereby motivating the need for a robust solution. Following this, we systematically analyze current approaches to robust zero-velocity detection, while categorizing public code and data. The article concludes with a discussion on commercialization along with guidance for future research.
\end{abstract}

\section{Introduction}

As described in classical mechanics, relative position is the integral of velocity. In a similar manner, relative position can also be computed as the double and triple integral of specific force and angular velocity, respectively. This is the basic idea of inertial navigation, where accelerometers and gyroscopes are used to measure specific force and angular velocity, respectively.

\begin{figure}[t]
	\def\svgwidth{3.5in}
	\hspace*{6mm}
	\scalebox{0.85}{
\begingroup%
  \makeatletter%
  \providecommand\color[2][]{%
    \errmessage{(Inkscape) Color is used for the text in Inkscape, but the package 'color.sty' is not loaded}%
    \renewcommand\color[2][]{}%
  }%
  \providecommand\transparent[1]{%
    \errmessage{(Inkscape) Transparency is used (non-zero) for the text in Inkscape, but the package 'transparent.sty' is not loaded}%
    \renewcommand\transparent[1]{}%
  }%
  \providecommand\rotatebox[2]{#2}%
  \newcommand*\fsize{\dimexpr\f@size pt\relax}%
  \newcommand*\lineheight[1]{\fontsize{\fsize}{#1\fsize}\selectfont}%
  \ifx\svgwidth\undefined%
    \setlength{\unitlength}{612.28346457bp}%
    \ifx\svgscale\undefined%
      \relax%
    \else%
      \setlength{\unitlength}{\unitlength * \real{\svgscale}}%
    \fi%
  \else%
    \setlength{\unitlength}{\svgwidth}%
  \fi%
  \global\let\svgwidth\undefined%
  \global\let\svgscale\undefined%
  \makeatother%
  \begin{picture}(1,0.64351852)%
    \lineheight{1}%
    \setlength\tabcolsep{0pt}%
    \put(0,0){\includegraphics[width=\unitlength,page=1]{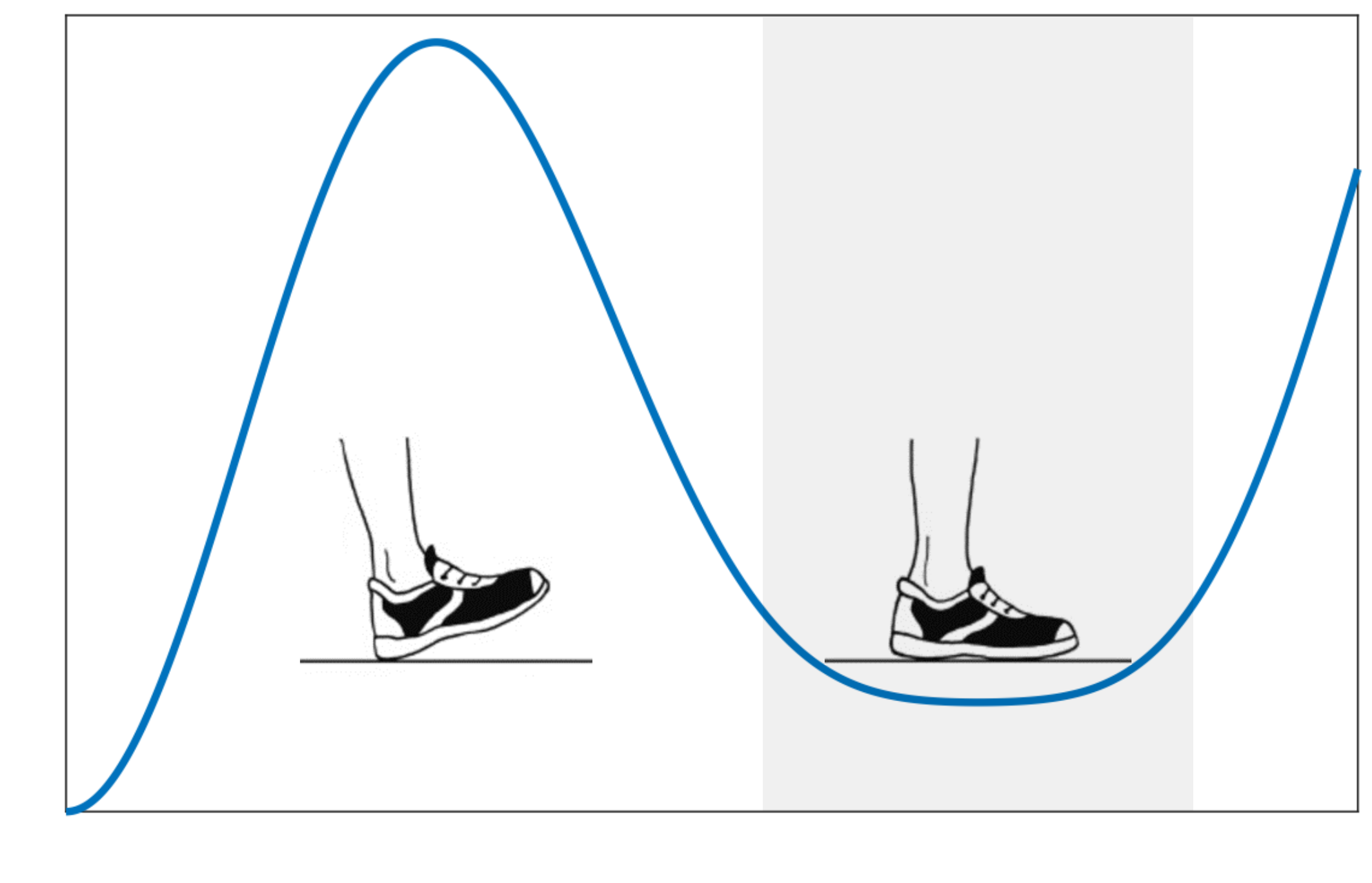}}%
    \put(0.19981222,0.34132887){\color[rgb]{0,0,0}\makebox(0,0)[lt]{\lineheight{1.25}\smash{\begin{tabular}[t]{l}Foot is moving\end{tabular}}}}%
    \put(0.56700431,0.3412814){\color[rgb]{0,0,0}\makebox(0,0)[lt]{\lineheight{1.25}\smash{\begin{tabular}[t]{l}Foot is stationary\end{tabular}}}}%
    \put(0.6549535,0.07417071){\color[rgb]{0,0,0}\makebox(0,0)[lt]{\lineheight{1.25}\smash{\begin{tabular}[t]{l}ZUPTs\end{tabular}}}}%
    \put(0.22588585,0.11341618){\color[rgb]{0,0,0}\makebox(0,0)[lt]{\lineheight{1.25}\smash{\begin{tabular}[t]{l}Stand-alone \end{tabular}}}}%
    \put(0.48055815,0.00613601){\color[rgb]{0,0,0}\makebox(0,0)[lt]{\lineheight{1.25}\smash{\begin{tabular}[t]{l}Time\end{tabular}}}}%
    \put(0.02457175,0.25440859){\color[rgb]{0,0,0}\rotatebox{90}{\makebox(0,0)[lt]{\lineheight{1.25}\smash{\begin{tabular}[t]{l}Position error\end{tabular}}}}}%
    \put(0.17458577,0.06691572){\color[rgb]{0,0,0}\makebox(0,0)[lt]{\lineheight{1.25}\smash{\begin{tabular}[t]{l}inertial navigation\end{tabular}}}}%
  \end{picture}%
\endgroup%
}
	\caption{Zero-velocity updates are used to reduce the error drift of foot-mounted inertial navigation systems.}
	\label{fig_intro}
\end{figure}
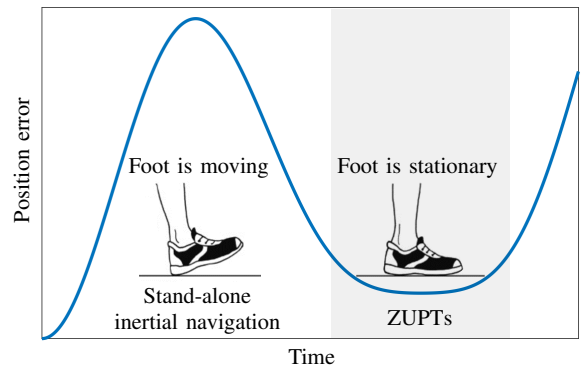

Following the rapid development of micro-electro-mechanical-systems (MEMS) technology, inertial sensors have found use in a diverse set of applications, including vibration monitoring, motion tracking, and road infrastructure monitoring \cite{Collin2019}. Thanks to their low cost, small size, and low weight, inertial sensors are also easy to incorporate into consumer devices, such as smartphones, tablets, and smartwatches \cite{Shaeffer2013}. However, despite their rapid proliferation, the limited performance of contemporary low-cost inertial sensors only enables accurate stand-alone inertial positioning over a very limited period of time  (typically only a few seconds). To enable practical low-cost inertial positioning in unexplored environments with no prior map information and no reliance on other sensors, the standard inertial navigation equations need to be complemented with  additional motion information. Next, we will discuss the most common way to do this within foot-mounted inertial navigation, that is, to incorporate zero-velocity updates (ZUPTs); see Fig. \ref{fig_intro}.

\section{A Brief History of ZUPTs in \\Foot-Mounted Inertial Navigation}

Currently, standard foot-mounted inertial navigation systems use Bayesian inference methods to combine i) the inertial navigation equations, used to integrate the inertial measurements and thereby provide position, velocity, and orientation estimates, ii) ZUPTs, used to correct these position, velocity, and orientation estimates and thereby reduce the error drift, and iii) a zero-velocity detector, used to identify the sampling instances where ZUPTs are to be applied\footnote{The focus of the present review is on foot-mounted inertial navigation solutions for unexplored environments. Thus, detailed discussions on how to incorporate maps or other sensors are omitted.} \cite{Nilsson2014}. A navigation system that resembles this framework in several ways was described back in the early 2000s\footnote{Prior to this, foot-mounted inertial navigation with ZUPTs had been described in an unpublished DARPA project from the mid 1990s \cite{Sher1996}, and in a SPIE conference article from 1999 \cite{Elwell1999}.} \cite{Sagawa2000}. In \cite{Sagawa2000}, the zero-velocity instances were detected using gyroscopes, and the roll and pitch angles during the stance phase were estimated using accelerometers. During the swing phase, the orientation was estimated by integrating the gyroscope measurements. These orientation estimates were then used to project the accelerometer measurements onto the horizontal plane, which, in turn, enabled estimation of the horizontal velocity and the horizontal walking distance.

Subsequently, in the first half the '00s, various ad hoc methods for detecting zero-velocity instances and for applying ZUPTs within foot-mounted inertial navigation appeared in the literature \cite{Stirling2003,Randell2003,Veltink2003,Cavallo2005,Scapellato2005,Pappas2001,Sabatini2005}. This included navigation systems utilizing empirical step-length estimation \cite{Stirling2003,Randell2003}, motion constraints assuming walking over a flat surface \cite{Veltink2003}, estimation of systematic sensor errors \cite{Veltink2003,Cavallo2005,Scapellato2005}, and gait cycle segmentation \cite{Pappas2001,Sabatini2005}. Moreover, there was also some brief discussion about the difficulty of step detection under varying gait speeds \cite{Sabatini2005}.

\subsection{The State-Space Model Formulation}

The first major technological leap came in 2005 with the introduction of the nonlinear state-space model~\cite{Foxlin2005} 
\begin{subequations}
\label{eq_state_space_model}
\begin{align}
    \label{eq_in_nav_eq}
    \mathbf{x}_{k+1}&=\mathbf{f}(\mathbf{x}_k,\mathbf{u}_k)+\mathbf{w}_k, \\[0.8ex]
    \mathbf{y}_k&=\mathbf{h}(\mathbf{x}_k)+\mathbf{e}_k,\quad \forall\hspace{0.15mm}k \text{ s.t. }D_{\boldsymbol{\theta}\hspace{-0.2mm},k}=1,
    \label{eq_zupt_model}
\end{align}
\end{subequations}
%
that combines the inertial navigation equations $\mathbf{f}(\cdot)$ with the zero-velocity measurement function $\mathbf{h}(\cdot)$, pseudo measurements of zero velocity $\mathbf{y}_k\triangleq\mathbf{0}$, and the zero-velocity detector $D_{\boldsymbol{\theta}\hspace{-0.2mm},k}\rightarrow\{0,1\}$. The zero-velocity measurement model is only applied at sampling instances where the zero-velocity detector takes the value 1, that is, when the sensor unit is considered stationary. The behavior of the zero-velocity detector is controlled by the design parameters $\boldsymbol{\theta}$. Furthermore, $\mathbf{x}_k$, $\mathbf{u}_k$, $\mathbf{w}_k$, 
and $\mathbf{e}_k$ denote the navigation state (incorporating three-dimensional position, velocity, and orientation, as well as possible auxiliary states for the modeling of systematic sensor errors), the inertial measurements, the inertial measurement errors, 
and the zero-velocity measurement errors, respectively. 
The errors $\mathbf{w}_k$ and $\mathbf{e}_k$ are typically assumed to be mutually uncorrelated zero-mean processes with known covariance matrices $\mathbf{Q}_k$ and $\mathbf{R}_k$, respectively. The subindex $k$ is used to denote quantities at sampling instance $k$.

On the one hand, it may be argued that the introduction of the nonlinear state-space model \eqref{eq_state_space_model} was a rather minor contribution since the same model had already been in use for several decades in vehicle applications \cite{Adams1979,Hadfield1980,Kelly1994}. Similarly, the idea of using body-worn inertial sensors to extract information on human motion had also been around for a long time \cite{Bassey1987,Willemsen1990,Jaarsveld1990}. Nevertheless, the revival of the state-space model formulation for foot-mounted inertial navigation offered several immediate benefits: i) it provided point estimates $\hat{\mathbf{x}}_k$ and associated estimation error covariance matrices $\mathbf{P}_k$ for three-dimensional position, velocity, and orientation based on a well-known and sound statistical framework \cite{Sarkka2013}, ii) it enabled a rigorous observability analysis \cite{Nilsson2013}, iii) it allowed for straightforward extensions to fusion with other sensor measurements or motion constraints in a probabilistic manner \cite{Zeng2017}, iv) it formalized the previous heuristic motion constraints in a simple, yet efficient manner\footnote{As expressed in \cite{Foxlin2005}, ``Introducing ZUPTs as measurements into the EKF [extended Kalman filter] instead of simply resetting the velocity to zero in the inertial integrator achieves important additional benefits. Most noticeably, the ZUPT lets the EKF retroactively correct most of the position drift that occurs during the stride phase. This is possible because the EKF tracks the growing correlations between the velocity and position errors in certain off-diagonal elements of the covariance matrix.''}, and v) it made it possible to solve the nonlinear state-space model \eqref{eq_state_space_model} by choosing among a multitude of established nonlinear filters and smoothers \cite{Zampella2012}. 

\begin{figure}[t]
	\def\svgwidth{3.5in}
	\hspace*{5mm}
	\scalebox{0.9}{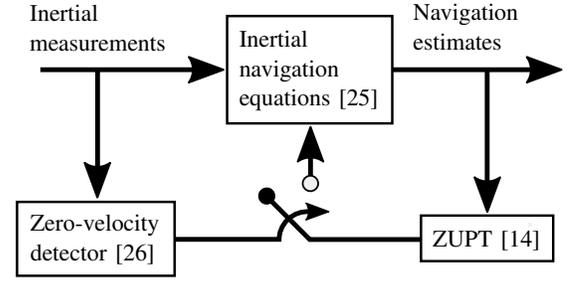}
	\caption{\label{figure_flowchart}The relationship between the inertial measurements, the inertial navigation equations, the zero-velocity detector, the ZUPTs, and the navigation estimates in the extended Kalman filter \protect\cite{Nilsson2012b}. When the zero-velocity detector detects a zero-velocity instance, a ZUPT, that is, a measurement update, is used to correct the navigation estimates.}
\end{figure}

\subsection{Formalization of Zero-Velocity Detection}
\label{section_formalization_zv_detection}

Once the state-space model had been established, the main question was how to design the zero-velocity detector $D_{\boldsymbol{\theta}\hspace{-0.2mm},k}$. Several heuristic detection methods were proposed. These included \emph{acceleration-moving variance detectors}, based on the moving variance of the accelerometer measurements \cite{Veltink1996,Kwakkel2008,Godha2006,Godha2008}; \emph{acceleration-magnitude detectors}, based on whether the magnitude of the accelerometer measurements was close to the gravitational acceleration \cite{Krach2008,Godha2006,Godha2008}; and \emph{angular rate energy detectors}, based on the energy of a window of gyroscope measurements \cite{Cavallo2005,Ojeda2007,Feliz2009}. The issue was settled when it was shown that all of these detectors can be derived within the same generalized likelihood-ratio test framework, given different prior knowledge about the sensor signals \cite{Skog2010}. In particular, let ${\cal H}_0$ and ${\cal H}_1$ denote the hypotheses that the sensor unit is moving and that the sensor unit is stationary, respectively. The likelihood-ratio test based on the measurements $\mathbf{z}_k\triangleq\{\mathbf{u}_n\}_{n=k-w_b}^{k+w_f}$, where $w_b$ and $w_f$ are some chosen backward and forward window lengths, then decides on hypothesis ${\cal H}_1$ if and only if
\begin{equation}
\label{eq_lrt_for_zv_detection}
L(\mathbf{z}_k)\triangleq\frac{p(\mathbf{z}_k|{\cal H}_1)}{p(\mathbf{z}_k|{\cal H}_0)}>\gamma
\end{equation}
where $\gamma$ is some user-specified threshold. The formulation of zero-velocity detection as a generalized likelihood-ratio test also motivated the use of the stance hypothesis optimal detection (SHOE) detector, which was derived as the generalized likelihood-ratio test resulting from using all available inertial sensors and all applicable motion models. Empirical results supported the use of the SHOE detector and indicated that the most valuable information for zero-velocity detection is contained in the gyroscope measurements \cite{Skog2010b,Olivares2012}. However, for the purpose of zero-velocity detection, the accelerometer and gyroscope measurements complement each other, and it will often be beneficial to use them both \cite{Callmer2010}. 

\begin{figure}[t]
	\hspace*{-2.5mm}
	\vspace*{0mm}
	\includegraphics{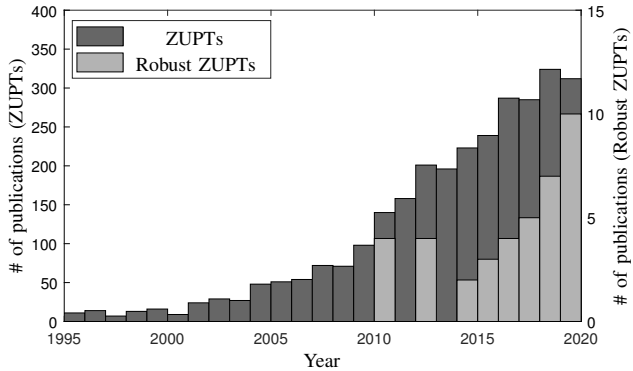}
	\caption{The number of publications including the term ``ZUPT'' (based on Google Scholar searches) and the number of articles on robust ZUPTs within foot-mounted inertial navigation.}
	\vspace*{0mm}
	\label{figure_bibliometrics}
\end{figure}

Generally, the likelihood-ratio test in \eqref{eq_lrt_for_zv_detection} is applied to overlapping windows $\{\dots,\mathbf{z}_{k-1},\mathbf{z}_k,\mathbf{z}_{k+1},,\dots\}$. The zero-velocity detector $D_{\boldsymbol{\theta}\hspace{-0.2mm},k}$ is then set to 1 at sampling instance $k$ if and only if at least one of the windows $\{\mathbf{z}_{k-w_f},\dots,\mathbf{z}_{k+w_b}\}$ (the windows including $\mathbf{u}_k$) results in a decision on hypothesis ${\cal H}_1$. The relationship between the inertial measurements, the inertial navigation equations, the zero-velocity detector, the ZUPTs, and the navigation estimates is illustrated in Fig. \ref{figure_flowchart}. The growing research interest in ZUPTs and foot-mounted inertial navigation is illustrated in Fig. \ref{figure_bibliometrics}. For further details on the implementation of a foot-mounted inertial navigation system, refer to \cite{Nilsson2014,Nilsson2012b}, and \cite{Fischer2013}. Having reviewed the standard state-space model and the most commonly used zero-velocity detectors, we are now ready to take a closer look at the sources of error within foot-mounted inertial navigation. 

\section{Sources of Error}
\label{section_error_sources}

A foot-mounted inertial navigation system computes navigation estimates by fusing information from sensor measurements and motion models. In practice, both the sensor measurements and the motion models are subject to imperfections that contribute to the final estimation error. In addition, interaction effects exist, so that imperfections in the motion models exacerbate the negative effect of sensor errors, and vice versa. In what follows, we will discuss the effect of both sensor errors and modeling errors. 

\subsection{Sensor Errors}
\label{section_sensor_errors}

Off-the-shelf inertial sensors commonly used for foot-mounted inertial navigation have several shortcomings, including random noise, systematic sensor errors, limited bandwidth, limited sampling rate, and limited dynamic range \cite{Groves2008}. The effect of random noise and sensor biases can be studied by analysing the observability properties of the state-space model \eqref{eq_state_space_model} when the sensor biases are included in the state vector. In this way, it can be shown that ZUPTs provide observability of all navigation states and all sensor biases except the position, the yaw angle, and the gyroscope bias in the direction of gravity \cite{Foxlin2005,Nilsson2013}. As a result, the errors in the position and yaw estimates will gradually increase over time. 
Refer to \cite{Wang2019c} and \cite{Nilsson2010} for studies on the relationship between sensor noise parameters and positioning performance.

During normal gait, the largest accelerations will occur during the heel strike, that is, when the heel hits the ground just prior to the stance phase. During the heel strike, the accelerations may exceed $15\,\text{g}$, with the main impulse lasting less than $3\,\text{ms}$ \cite{Ju2015}. As a result of limitations in bandwidth, sampling rate, and dynamic range, the accelerometers may not be able to correctly measure these extreme accelerations \cite{Nilsson2012a}. This tends to lead to estimation errors, particularly along the vertical axis. One way to counteract this is to use shock absorbing pads \cite{Li2012}. Additionally, it is possible to modify the noise or estimation error covariance matrices during or after the heel strike so that they better reflect the true system uncertainties. This can be done, for example, by setting all covariances in $\mathbf{P}$ involving the velocity estimates to zero during the stance phase \cite{Ju2015}, or by increasing elements in the sensor noise covariance matrix $\mathbf{Q}$ when a heel strike is detected \cite{Ju2018a}. The negative effects during the heel strike can also be mitigated by sensible choices of sensor placement. One option is to place the inertial sensors on the heel so that the assumption of zero velocity can be utilized already at the heel strike \cite{Ju2015}. Another option is to place additional inertial sensors on the calf, since the calf does not experience such extreme accelerations as the foot \cite{Ju2018a}. 

\begin{table*}[t]
	\small
	\caption{Some of the error sources within foot-mounted inertial navigation. \label{table_error_sources}}
	\centering
	\begin{tabular}{l|l|l|l|l}
		\hline
		\hline
		& \multicolumn{2}{c|}{} & \multicolumn{2}{c}{} \\ [-2.1ex]
	    & \multicolumn{2}{c|}{Sensor errors} & \multicolumn{2}{c}{Modeling errors} \\
	    & \multicolumn{2}{c|}{Section \ref{section_sensor_errors}} & \multicolumn{2}{c}{Section \ref{section_modeling_errors}} \\
		& \multicolumn{2}{c}{} & \multicolumn{2}{c}{} \\ [-2.5ex]
		\hline    
		& & & \multicolumn{2}{c}{} \\ [-2.1ex]
		Error source & Random noise and & Limited sampling rate, & \multicolumn{2}{l}{Non-zero mean of ZUPT errors, temporal}  \\
		& systematic sensor errors & bandwidth, dynamic range & \multicolumn{2}{l}{correlation of ZUPT errors} \\
		& & & \multicolumn{2}{c}{} \\ [-2.5ex]
		\hline    
		& & & \multicolumn{2}{c}{} \\ [-2.1ex]
		Gait cycle phase$^{\text{a}}$ & Swing phase & Heel strike & \multicolumn{2}{c}{Zero-velocity instances} \\
		& & & \multicolumn{2}{c}{} \\ [-2.5ex]
		\hline
		& & & \multicolumn{2}{c}{} \\ [-2.1ex]
		Potential & Position and yaw drift & Systematic errors along & \multicolumn{2}{l}{Stride length underestimation \cite{Peruzzi2011}, systematic} \\
	    consequences & \cite{Foxlin2005,Nilsson2013,Wang2019c,Nilsson2010} & the vertical axis \cite{Ju2015} & \multicolumn{2}{l}{errors along the vertical axis \cite{Nilsson2012a}} \\
		& & & \multicolumn{2}{l}{} \\ [-2.5ex]
		\hline
		& & & \multicolumn{2}{l}{} \\ [-2.1ex]
		Suggested & Better sensors \cite{Foxlin2005}, & Adjust error covariances & \multicolumn{2}{l}{Mount sensors close to the heel \cite{Peruzzi2011}, use} \\
	    remedies & inertial arrays \cite{Wahlstrom2018,Skog2014}, & \cite{Ju2015}, place additional & \multicolumn{2}{l}{non-zero velocity updates \cite{Wu2014}, estimate} \\
        & sensors on both feet \cite{Zhao2019b} & sensors on calf \cite{Ju2018a}, use & \multicolumn{2}{l}{vertical position using the pitch of the foot} \\
        & & shock absorbing pads \cite{Li2012} & \multicolumn{2}{l}{at midstance \cite{Park2011}} \\
		\multicolumn{5}{c}{}  \\ [-2.4ex]
		\hline \hline 
		\multicolumn{5}{c}{}  \\ [-2.2ex]
		\multicolumn{5}{l}{\footnotesize{$^{\text{a}}$ Refers to the phase of the gait cycle when the errors have their greatest effect on the navigation system.}} \\ [-0.2ex]
	\end{tabular}
	\vspace*{-0mm}
\end{table*}

It should be emphasized that analytical analyses of observability or error propagation (see \cite{Foxlin2005,Nilsson2013,Wang2019c}) typically do not consider modeling errors in the ZUPTs or limitations in bandwidth, sampling rate, or dynamic range. This means that while the observability analyses are correct per se, their utility is diminished by the limited validity of the considered state-space models. Hence, it is important to be careful when drawing conclusions from such analyses. As an example, a straightforward observability analysis shows that all sensor biases except the gyroscope bias in the direction of gravity are made observable by ZUPTs \cite{Foxlin2005}. However, in practice, environment- and gait-dependent errors, such as modeling errors or errors due to limitations in bandwidth, may have a significant impact on the dead-reckoning errors. Therefore, the choice of estimating biases without accounting for modeling errors may worsen performance in a scenario with varying environment and gait conditions \cite{Nilsson2012a,Ju2015}. 

Rather than attempting to estimate the gyroscope bias using all available measurements, a better approach is to apply so-called zero-angular-rate updates (ZARU) or zero-integrated-heading rate (ZIHR) updates at detected  zero-angular-rate events \cite{Ashkar2013}. Generally, ZARUs should only be applied when the angular velocities are significantly smaller than the gyroscope bias. This means that zero-angular-rate events cannot be detected using conventional zero-velocity detectors, such as the SHOE detector and the angular rate energy detector, which are sensitive to the impact of the gyroscope bias\footnote{Norm-based detectors, such as the SHOE detector and the angular rate energy detector, cannot separate the gyroscope bias from other contributions to the norm. This is generally not an issue when performing zero-velocity detection, since the contribution of the gyroscope bias to the gyroscope norm during the stance phase tends to be small in comparison to the systematic angular velocities (see Section \ref{section_modeling_errors}). However, it is an issue when attempting to detect zero-angular-rate events. Refer to \cite{Shaolei2018} for experimental comparisons of variance- and norm-based detectors.}. Instead, zero-angular-rate events should be detected using the temporal variance of the gyroscope measurements, which is unaffected by the gyroscope bias. In comparison to ZUPTs, the utility of ZARUs is rather limited since systematic angular velocities during the stance phase often prevent ZARUs from being applied during normal gait \cite{Abdulrahim2012,Bancroft2012}. Finally, note that inertial measurement units may come with built-in pre-filters that can influence the navigation system in a negative way. For example, if the raw gyroscope and accelerometer measurements are subject to \emph{different} low-pass filters, this may lead to phase shifts which appear as a relative timing error between the gyroscope and accelerometer measurements \cite{Haidekker2013}. Information about how these filters are designed is typically not available to end-users of the sensor units.

In summary, the sensor measurements introduce significant errors into the navigation system. Therefore, ZUPTs are needed to break the cubic position error growth of stand-alone inertial navigation. However, while ZUPTs reduce the impact of the sensor errors, they also introduce several modeling errors. In the following subsection, we examine the zero-velocity modeling errors in more detail.

\subsection{Modeling Errors}
\label{section_modeling_errors}

There are two sources of modeling errors in foot-mounted inertial navigation: the inertial navigation equations in \eqref{eq_in_nav_eq}, and the zero-velocity model in \eqref{eq_zupt_model}. The continuous-time\footnote{In practice, the inertial navigation equations need to be discretized due to the limited sampling rate of inertial sensors. The effect of the sampling rate on the estimation accuracy is discussed in Section \ref{section_sensor_errors}.} inertial navigation equations are perfect in the sense that, given perfect knowledge of the initial navigation state and of the specific force and angular velocity at all time instances, they will produce perfect navigation estimates \cite{Groves2008}. The zero-velocity model, on the other hand, make several implicit assumptions that have been shown not to hold in typical usage scenarios. In particular, commonly applied filters and smoothers only guarantee optimality\footnote{Strictly speaking, the commonly applied extended Kalman framework only provides an optimal solution to a linearization of the true model and is not optimal in terms of mean-square error.} if the zero-velocity measurement errors $\mathbf{e}$ are zero-mean and white \cite{Sarkka2013}. However, as has been demonstrated in several studies, the foot will typically undergo systematic motions during the stance phase, and thereby violate the zero-velocity assumption. This is particularly evident at high gait speeds \cite{Bebek2010,Peruzzi2011,Nilsson2012a,Ju2015}. As a result, the zero-velocity measurement errors $\mathbf{e}$ are not zero-mean and exhibit both intra- and inter-step correlations.

Evaluations using camera tracking systems have indicated that the minimum velocity of the foot during ordinary walking is of the order of $0.005\,\text{m}/\text{s}$ in the walking direction. Consequently, the conventional ZUPT has been estimated to result in an underestimation of the stride length by up to $0.7\%$ \cite{Peruzzi2011}. Although the minimum velocity of the foot increases with gait speed, the accompanying negative effect on the relative stride length error can be expected to be counterbalanced by the simultaneous decrease in the stance duration and increase of the stride length. To minimize the velocity during the stance phase, it has been suggested that the inertial sensors should be mounted close to the heel \cite{Peruzzi2011}. However, a later study claimed that mounting the sensors on the forefoot will give better positioning performance \cite{Wang2019b}. Instead of trying to adjust the dynamics, the errors of the zero-velocity model can also be reduced by using a more realistic model. One way to do this is to use the gyroscope measurements and the moment arm between the center of rotation and the sensor position to estimate the non-zero velocity at midstance \cite{Wu2014,Ju2018b}. The estimated velocity can then be used to apply \emph{non}-zero velocity updates. 

In addition to causing biased stride length estimates, violations of the zero-velocity model assumptions can also lead to other estimation errors. Since ZUPTs are often applied even though the sensor unit is not completely stationary, the steps may, in a sense, be “cut short” at both ends. In this case, motions from specific phases of the gait cycle will be systematically suppressed by the ZUPTs, which, in turn, tends to lead to systematic errors along the vertical axis \cite{Nilsson2012a}. One study proposed to improve the vertical position estimates by estimating the inclination of the ground based on the pitch of the foot at midstance, and then using the estimated ground inclination together with the estimated step length to correct the vertical position estimates \cite{Park2011}. However, this idea assumes that the pitch of the foot at midstance is a good indicator of the change in vertical position over a given step, and the method can therefore not be used when, for example, walking up or down stairs. Another source of modeling errors is the potential non-optimal tuning of the navigation system. The tuning of the measurement error covariance matrix $\mathbf{R}_k$, for example, is made difficult by the fact that the standard deviation will be dependent on many factors, including the sensor placement, the walking surface, and the design of the zero-velocity detector \cite{Wang2019b}. The parameter tuning is further complicated by the violation of several inference assumptions, for example, the previously described modeling errors in the zero-velocity model. In conclusion, the modeling errors in the zero-velocity model originate from a large number of sources, and are heavily affected by environment and gait conditions. Research on how to reduce these modeling errors is currently scarce.

Table \ref{table_error_sources} lists some of the most important sources of error, together with potential consequences for the navigation solution, remedies suggested in the literature, and the phase of the gait cycle when the errors are introduced into the navigation system. Due to the interdependent effect of sensor and modeling errors, great caution should be exercised when interpreting the table. In practice, it may not be possible to separate the impact of different error sources; therefore, the table should only be seen as a simplified summary, created to aid in the understanding of some of the main error sources.

\section{Why Is There No One Threshold \\to Rule Them All?}\label{section_rule_them_all}

The description of error sources in the preceding section will now 
provide the foundation for a discussion of the performance characteristics of the foot-mounted inertial navigation solution. The navigation performance is dependent on many factors. A slower gait, for example, reduces the impact of both sensor errors and modeling errors. The less extreme dynamics at the heel strike reduce the negative effect of sensor limitations in bandwidth, sampling rate, and dynamic range, and the modeling errors are smaller since the true velocity when applying ZUPTs is closer to zero. Thus, in comparison to fast gait, modest gait speeds generally result in both better navigation performance and lower sensitivity to parameter settings \cite{Wahlstrom2019,Wahlstrom2020,Skog2010b}. Likewise, the performance generally improves with higher sensor quality \cite{Collin2019}, higher sampling rates \cite{Wang2018b}, sensor placements on the forefoot or close to the heel \cite{Wang2019b,Peruzzi2011}, and hard walking surfaces \cite{Wang2019b}.

\begin{figure}[t]
	\hspace*{0mm}
	{\hspace*{-7mm}
	\includegraphics{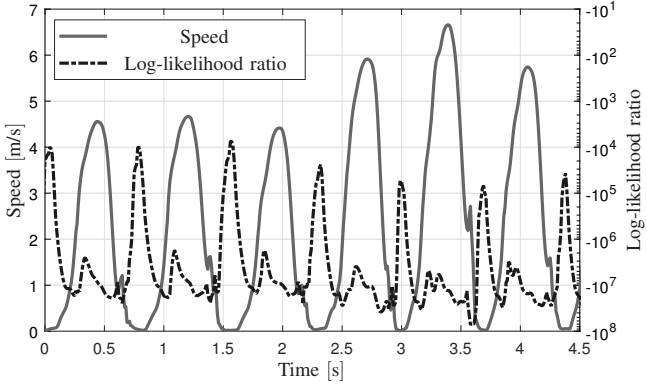}
	}
	\caption{The relationship between gait speed and log-likelihood ratio for zero-velocity detection. The data was taken from ID 15 in \protect\cite{Angermann2010}.}
	\vspace*{0mm}
	\label{fig_speed_vs_loglikelihood}
\end{figure} 

\begin{figure}[t]
	\vspace*{-0.2mm}
	\hspace*{-1mm}
	{
		\vspace*{0mm}
		\includegraphics{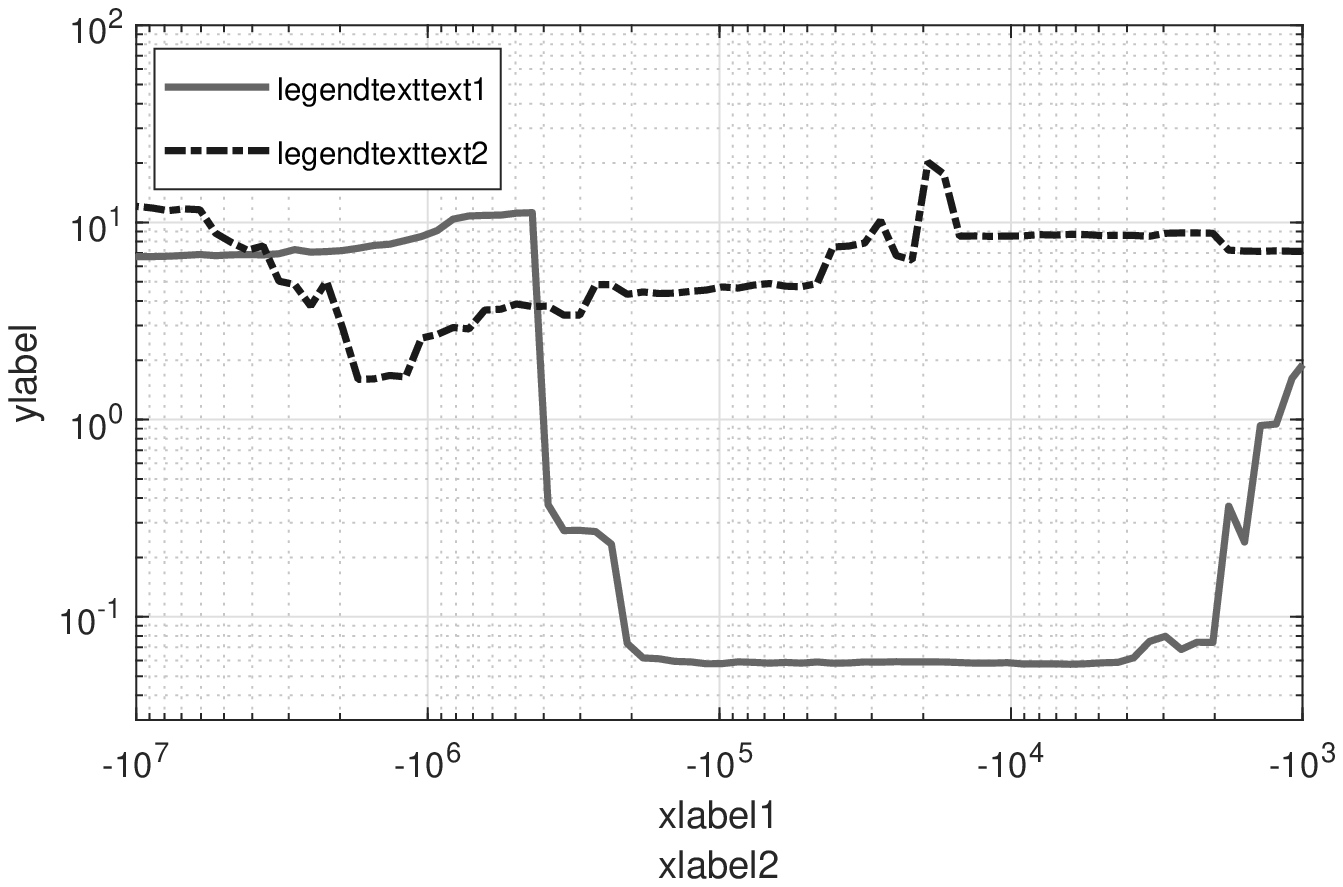}
	}
	\hspace*{-1mm}
	{
		\vspace*{0mm}
		\includegraphics{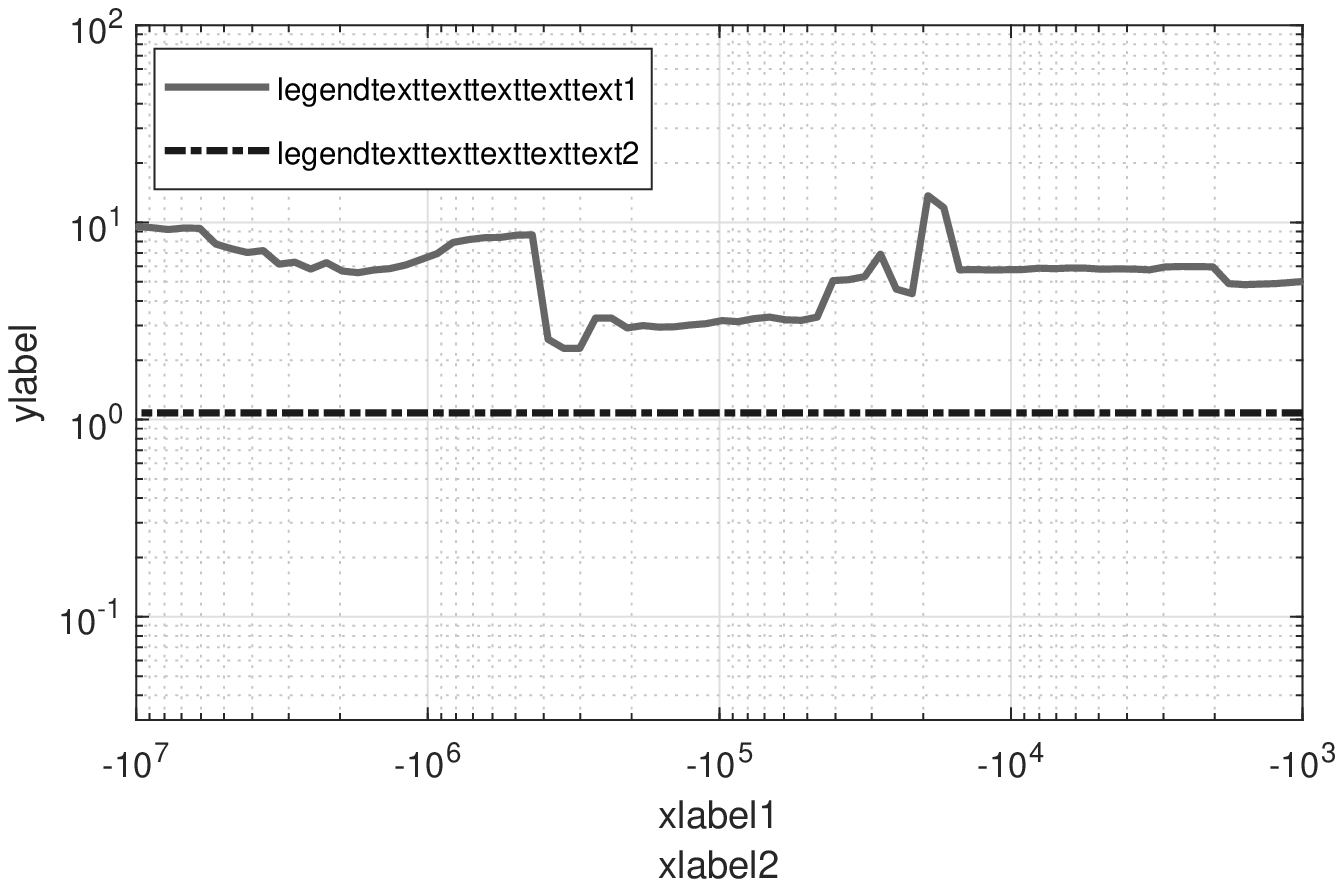}
	}
	\vspace*{-5mm}
	\caption{The relationship between gait speed, zero-velocity detection threshold, and positioning performance, illustrated using data from \protect\cite{Angermann2010}. The walking and running data was taken from IDs 3 and 7, respectively. In (a), the RMSE was computed separately for data with walking and running. In (b), the RMSE was computed using all  available data. In both (a) and (b), the position estimates were computed using an extended Kalman smoother \cite{Colomar2012} implemented with the SHOE detector \cite{Skog2010}.
	}
	\vspace*{0mm}
	\label{fig_rmse_angermann}
\end{figure}  

One issue that complicates the implementation of foot-mounted inertial navigation systems is the interdependent effect of parameter settings and environment- or gait-dependent factors on navigation performance. In other words, to achieve optimal performance, system parameters must be continuously adapted to both the environment and the gait style, both of which may be rapidly changing. To make this more concrete, we will consider the relationship between the zero-velocity detection threshold. i.e., $\gamma$ in \eqref{eq_lrt_for_zv_detection}, and the gait speed. Since the sensor errors that accumulate during stand-alone inertial navigation eventually always lead to large estimation errors, it is important that the navigation system does not run for too long without any ZUPTs. However, if too many ZUPTs are applied, or if ZUPTs are applied at points in time when the true velocity is far from zero, the accompanying modeling errors will reduce the navigation performance considerably. In other words, ZUPTs reduce the impact of sensor errors at the expense of introducing modeling errors, and the zero-velocity detection threshold can be said to represent the trade-off, calibrated for a specific set of environment and gait conditions, between the sensor errors and the modeling errors. Therefore, in the situation illustrated in Fig. \ref{fig_speed_vs_loglikelihood}, where the user first walks and then runs, the calibration of the threshold is notably difficult. In particular, note how the values of the likelihood ratio during the stance phase decrease when the pedestrian starts running after about $2.3\,\text{s}$. If the threshold is too large, an insufficient number of ZUPTs will be applied when the user is running. Likewise, if the threshold is too small, too many ZUPTs will be applied when the user is walking. 

\begin{figure*}[t]
	\def\svgwidth{7.5in}
	\hspace*{2mm}
	\scalebox{0.9}{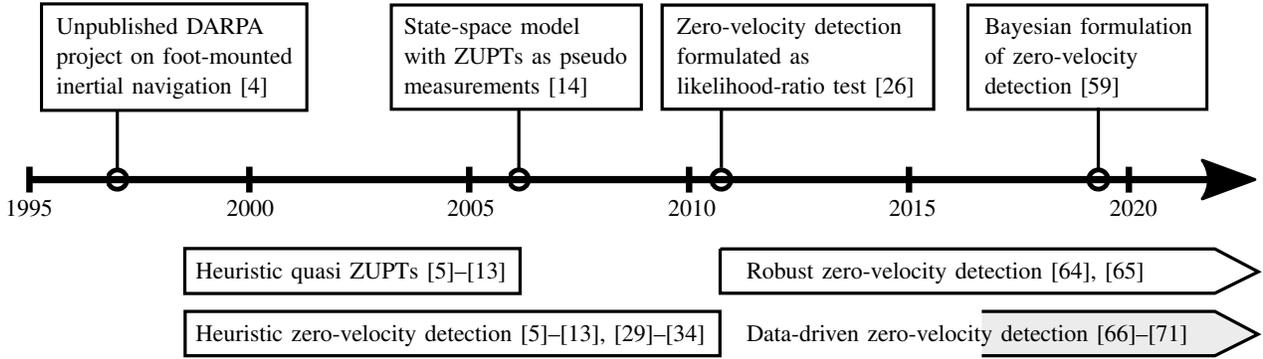}
	\caption{\label{figure_timeline}Timeline of ZUPTs in foot-mounted inertial navigation.}
\end{figure*}

\begin{figure*}[t]
	\def\svgwidth{7.5in}
	\hspace*{1mm}
	\scalebox{0.9}{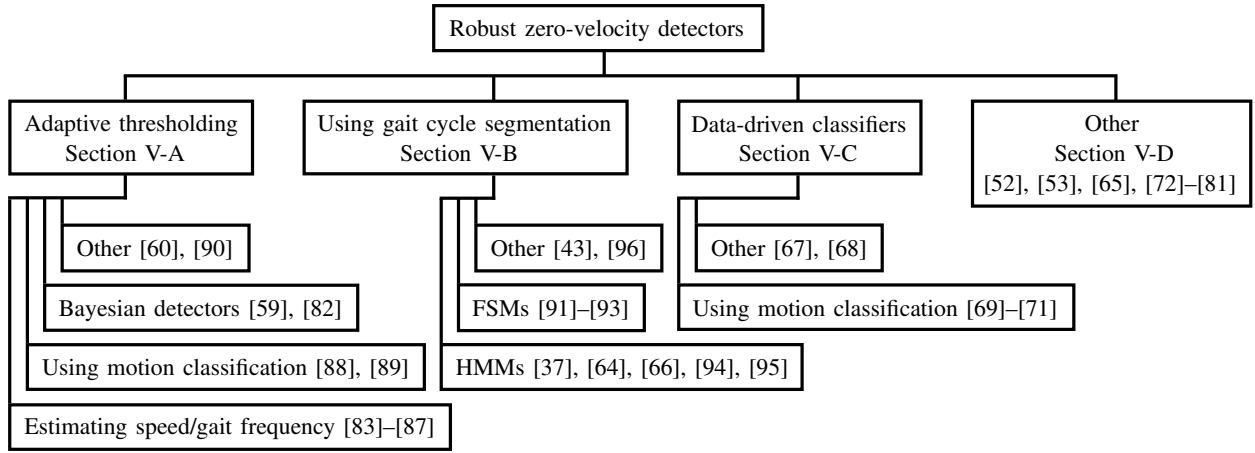}
	\caption{\label{figure_hierarchical_classification}Hierarchical classification tree for robust zero-velocity detectors.}
\end{figure*}

To further illustrate this issue, Fig. \ref{fig_rmse_angermann} (a) displays the time-averaged position root-mean-square error (RMSE) as dependent on the zero-velocity detection threshold when walking and running. For each gait speed, about 30 seconds of data was considered. As discussed earlier, a slower gait speed is seen to give both better performance and lower parameter sensitivity. Moreover, as expected given the relationship between gait speed and likelihood ratio, the optimal threshold for running is considerably lower than the optimal threshold for walking\footnote{It should be emphasized that, although the exact values for the likelihood ratio and the optimal threshold will be heavily dependent on both other system parameters and the sensor placement, the overall characteristics of their relationship to e.g., the gait speed will remain fixed across a broad range of different setups.}. Now, consider the problem of tuning the threshold so that the navigation system performs well for both gait speeds. A natural optimization function is the position RMSE, time-averaged over both data sets. This position RMSE is shown in Fig. \ref{fig_rmse_angermann} (b). As can be seen, the optimal threshold when considering all data is markedly different from the optimal threshold when only considering data from either of the two gait speeds. Thus, the best performance is obtained by adapting the threshold based on the gait speed. In fact, as illustrated in Fig. \ref{fig_rmse_angermann}, an adaptive navigation system, which alternates between the optimal thresholds for walking and running, reduces the position RMSE by more than 50\% in comparison with the best fixed threshold. In the next section, we will review methods for designing zero-velocity detectors that can adapt to the environment and gait style. 

\section{Robust Zero-Velocity Detection}
\label{section_adaptive_zupts}

To construct a zero-velocity detector that is robust with respect to variations in environment and gait style, the basic detection framework described in Section \ref{section_formalization_zv_detection} must be augmented or replaced. At first, research on robust zero-velocity detection made use of models describing, for example, the gait cycle or the time length of a typical step. In recent years, however, much of the focus has shifted to robust zero-velocity  detection using data-driven methods. These require a training phase, where data from high-accuracy, infrastructure-dependent sensors or similar is used to learn the zero-velocity detector. The development from the first studies on foot-mounted inertial navigation to recently presented robust zero-velocity detectors is illustrated in Fig. \ref{figure_timeline}. Fig. \ref{figure_hierarchical_classification} summarizes the robust zero-velocity detectors in a hierarchical classification tree.

\subsection{Detectors using Adaptive Thresholding}
\label{section_threshold_adjustment}

\begin{figure}[t]
	\def\svgwidth{3.5in}
	\hspace*{0.75mm}
	\scalebox{0.9}{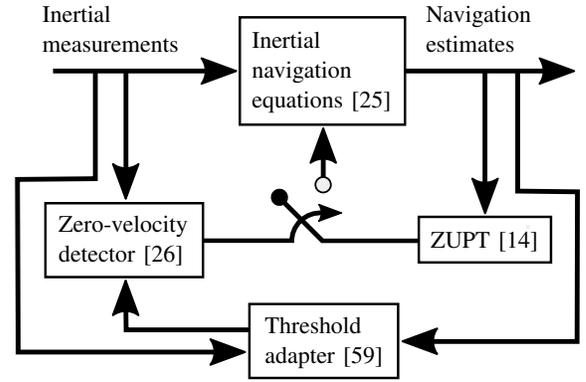}
	\caption{\label{figure_flowchart_adaptive}A foot-mounted inertial navigation system with a threshold adapter. Compare with Fig. \ref{figure_flowchart}.}
\end{figure}

As illustrated in Fig. \ref{figure_flowchart_adaptive}, a common approach to robust zero-velocity detection is to use the overall detection framework described in Section \ref{section_formalization_zv_detection} and adapt the threshold based on changes in environment or gait conditions. In this case, the threshold $\gamma_k$ on $L(\mathbf{z}_k)$ in \eqref{eq_lrt_for_zv_detection} is set based on a mapping\footnote{In practice, the function \eqref{eq_threshold_mapping_function} could also consider inertial measurements obtained after sampling instance $k$.}
\begin{equation}
 \gamma_k=g(\mathbf{u}_{1:k},\hat{\mathbf{x}}_{1:k}),
 \label{eq_threshold_mapping_function}
\end{equation}
that depends on the inertial measurements $\mathbf{u}$ and the navigation state estimates $\hat{\mathbf{x}}$. One way to construct this mapping is to formulate the detection problem in a Bayesian setting. This means that the detector decides on hypothesis ${\cal H}_1$ if and only if
\begin{equation}
\frac{p({\cal H}_1|\mathbf{z}_k)}{p({\cal H}_0|\mathbf{z}_k)}>\eta
    \label{eq_bayesian_test_for_zv_detection}
\end{equation}
where $\eta$ is a loss factor quantifying the cost of incorrect decisions. The Bayesian zero-velocity detector in \eqref{eq_bayesian_test_for_zv_detection} can be shown to be equivalent to the likelihood ratio test in \eqref{eq_lrt_for_zv_detection} with the threshold 
\begin{equation}
\gamma\overset{_\Delta}{=}\frac{1-p({\cal H}_1)}{p({\cal H}_1)}\cdot\eta.
\end{equation}
Thus, the Bayesian formulation provides both a theoretical justification for adaptive zero-velocity detection within the established likelihood-ratio framework, and a sound way to design adaptive thresholds by modeling a time-varying hypothesis prior $p({\cal H}_1)$ and loss factor $\eta$. For example, if the loss factor is modeled so that it decays with the time since the last zero-velocity instance, the resulting adaptive threshold can be designed to both prevent the navigation system from running for too long without a ZUPT, and mitigate the risk of applying an excessive number of ZUPTs \cite{Wahlstrom2019}. Furthermore, one study proposed to adapt the parameter describing the loss factor decay based on the dynamics during the heel strike \cite{Wang2019a}.

Another way to design the mapping in \eqref{eq_threshold_mapping_function} is to first extract some relevant features from the inertial measurements and the navigation state estimates, and then learn a mapping from these features to the threshold using ground truth data. For example, if inertial measurements are used to train a supervised classifier to differentiate between a number of predefined motion classes, the optimal threshold for each motion class can then be found using a high-accuracy camera tracking system \cite{Wagstaff2017,Rantanen2018}. Similarly, the inertial measurements may be used to compute estimates of speed \cite{Zhang2017,Ma2017,Kim2020,Bai2020}, or gait frequency \cite{Tian2016}. These estimates can then be mapped to optimal threshold values by, for example, using ground truth speed from a treadmill \cite{Zhang2017}, by minimizing the closed-loop position error \cite{Ma2017}, or by minimizing the travelled distance error \cite{Bai2020}. Although most studies extract these features from foot-mounted inertial measurements and related estimates, there have also been proposals to utilize measurements from additional, chest-mounted accelerometer units \cite{Zhang2017}. 

One problem with zero-velocity detectors using adaptive thresholding is that a proper calibration of the mapping in \eqref{eq_threshold_mapping_function} tends to require large amounts of high-accuracy ground truth data. Since such data is not available in the unexplored environments where foot-mounted inertial navigation systems are typically meant to be used, this means that the adaptive detectors require an extensive calibration period, separate from the real-world deployment for which the navigation system is intended. Moreover, it also means that the mapping, once set, cannot adapt to real-time changes in gait or environment conditions that were not accounted for during the calibration process. One solution to these problems is to make use of the FootSLAM algorithm to transform the inertial odometry into position estimates with long-term error stability. These position estimates can then be used as pseudo ground truth in the threshold calibration. In this way, it is possible to design an adaptive zero-velocity detector that does not require map information, measurements from supplementary sensors, or user input \cite{Wahlstrom2020}. However, it should be noted that the method requires a period of time when the FootSLAM algorithm converges\footnote{In many cases, physical constraints provided by walls or other obstacles enforce sufficient consistency in the walking patterns to ensure convergence; refer to \cite{Angermann2012} for details.}. In addition, the pseudo ground truth provided by FootSLAM is limited by the fact that the FootSLAM algorithm cannot detect scale errors in the odometry. An alternative approach to reduce the need for calibration using ground truth data, rather than trying to account for all variations in environment and gait style in a ground truth-dependent training phase, is to examine relevant variations through their impact on the likelihood ratio over a typical gait cycle; see e.g., Fig. \ref{fig_speed_vs_loglikelihood}. Thus, after incorporating a step detector, the threshold can continuously be adjusted based on the likelihood ratio computed during previous steps \cite{Liu2014}.  
\subsection{Detectors using Gait Cycle Segmentation}
\label{section_gait_cycle}

\begin{figure}[t]
	\hspace*{-0mm}
	\vspace*{0mm}
	\includegraphics{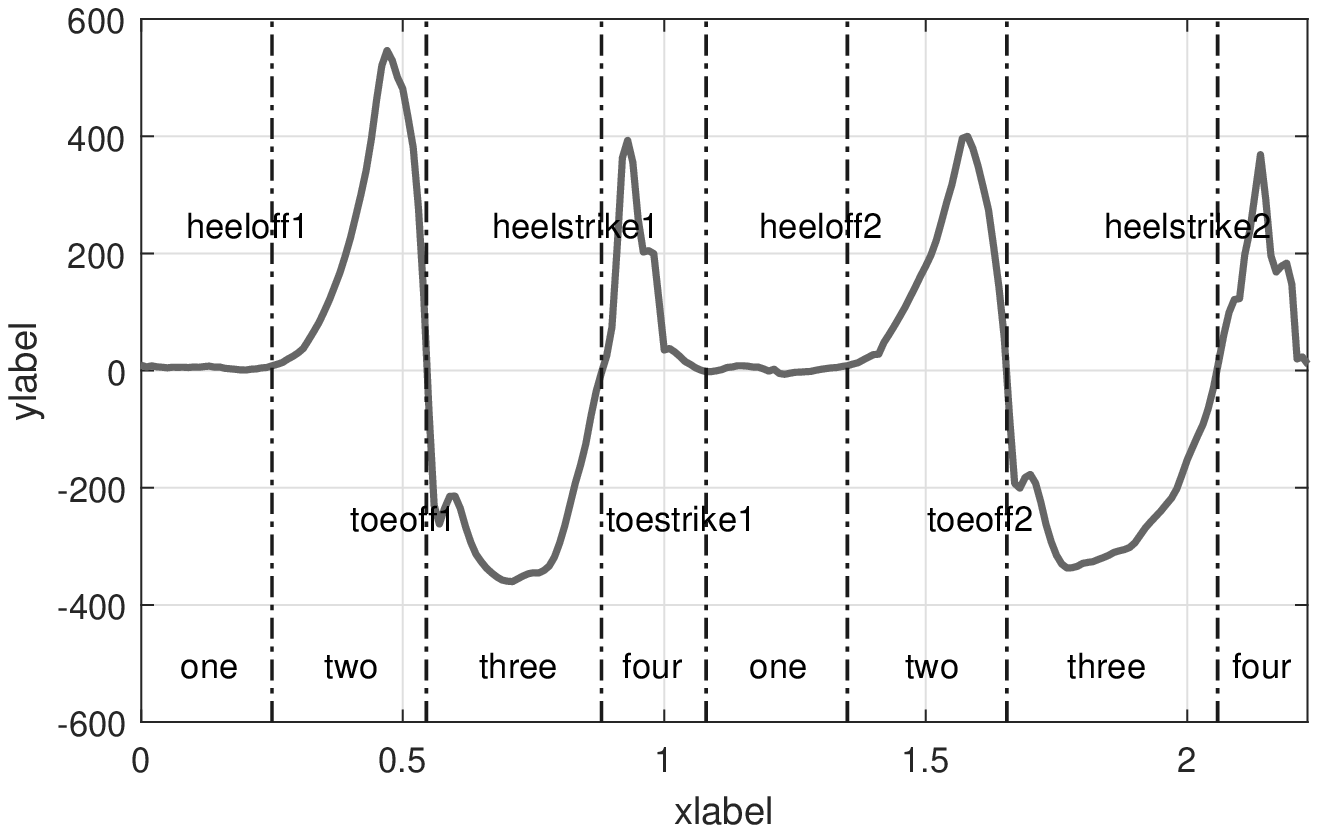}
	\caption{A gait cycle model used in several robust zero-velocity detectors \protect\cite{Park2010,Xu2015,Zhang2017b,Sun2018}. The data was taken from ID 15 in \protect\cite{Angermann2010}.}
	\vspace*{0mm}
	\label{fig_gait_cycle_segmentation}
\end{figure}

Several zero-velocity detectors attempt to increase robustness by exploiting characteristics in the gait cycle that in one way or another are invariant to changes in environment and gait style. Generally, these detectors use a model in which phases in the gait cycle are represented by a number of discrete states. As illustrated in Fig. \ref{fig_gait_cycle_segmentation}, the standard model includes four states, with the boundaries between two adjacent phases roughly given by the heel off, toe off, heel strike, and toe strike events \cite{Park2010,Xu2015,Zhang2017b,Sun2018}. That being said, several other gait cycle models have also been proposed \cite{Li2012,Ren2016}, including up to six states to model a conventional gait cycle \cite{Zhao2019a}, and up to nine states to also model backwards gait and walking up or down stairs. The state inference primarily uses gyroscope measurements in the lateral direction of the foot (one study instead used speed estimates \cite{Ren2016}), and can be based on hidden Markov models (HMMs) \cite{Park2010,Zhang2017b,Sun2018,Zhao2019a,Callmer2010}, finite-state machines (FSM) \cite{Ren2016,Yun2012,Ruppelt2016}, step detectors \cite{Li2012}, or heuristic rule-based methods \cite{Xu2015}. In the HMMs, the measurement function modeling the relationship between the discrete states and the measurements has been constructed using both neural networks \cite{Zhao2019a}, and Gaussian mixture models \cite{Sun2018}. In some models, one of the states corresponds to zero-velocity instances (for example, the state defined as the phase in between the toe strike and heel off events), and thus, the state inference immediately gives the zero-velocity instances being sought \cite{Sun2018,Ren2016,Callmer2010,Ruppelt2016}. However, in other models, the estimated state sequence is used as input to a zero-velocity detector that also considers other information, such as speed estimates \cite{Zhang2017b}, the results of conventional zero-velocity detectors \cite{Li2012,Xu2015}, or the time length of the detected zero-velocity instances \cite{Park2010}. Although there have also been evaluations considering walking up or down stairs or walking uphill/downhill \cite{Ren2016}, the experimental evaluations have mainly focused on adaptation to gait speed. The main limitation of zero-velocity detectors utilizing gait cycle segmentation is that it may be difficult to specify a gait cycle model that is general enough. While gait cycle models \emph{can} make the detector more robust, they can also worsen performance if the user's movements strongly deviate from the model assumptions.

\subsection{Detectors using Data-Driven Classifiers}
\label{section_data_driven_detectors}

Recently, robust zero-velocity detectors have been developed based on data-driven binary classifiers. In this case, the detection framework presented in Section \ref{section_formalization_zv_detection} is cast aside, and the decision of whether or not a zero-velocity instance has occurred is made directly based on the inertial measurements and the navigation estimates. In other words, the zero-velocity detector is defined as $D_{\boldsymbol{\theta}\hspace{-0.2mm},k}\triangleq T_{\boldsymbol{\theta}}(\mathbf{u}_{1:k},\hat{\mathbf{x}}_{1:k})$, where\footnote{In practice, the function \eqref{eq_data_driven_classifier} could also consider inertial measurements obtained after sampling instance $k$.}
\begin{equation}
\label{eq_data_driven_classifier}
    T_{\boldsymbol{\theta}}(\mathbf{u}_{1:k},\hat{\mathbf{x}}_{1:k})\rightarrow \{0,1\}
\end{equation}
is constructed by training against ground truth data. If the training data set is diverse enough, the detector can learn to perform well under highly varying conditions. The inference has been based on support vector machines \cite{Park2016}, random forests \cite{Kone2020}, histogram-based gradient boosting \cite{Kone2020}, long short-term memory (LSTM) neural networks \cite{Wagstaff2018,Zhu2019}, and convolutional neural networks \cite{Yu2019}. In addition to adapting to the environment and gait style through the input $\mathbf{u}_{1:k}$ and $\hat{\mathbf{x}}_{1:k}$, the detector can be made even more robust by subdividing the inference based on the results of a preceding motion classification \cite{Zhu2019,Kone2020}. For example, in \cite{Park2016}, an individual parameter set $\boldsymbol{\theta}$ is trained for 24 different motion classes. During the test phase, the foot-mounted inertial measurements are first used to classify the motion into one of these 24 motion classes. The zero-velocity detection is then performed using the parameter set $\boldsymbol{\theta}$ that was trained for that specific motion class.

Ground truth data for training the classifier in \eqref{eq_data_driven_classifier} has been obtained from camera tracking systems \cite{Zhao2019a,Kone2020}, ultrasonic ranging of the distance between the shoe and the floor \cite{Zhu2019}, and manual annotation of inertial measurements \cite{Yu2019}. Note that even if high-accuracy measurements of velocity are available, the creation of ground truth zero-velocity labels is still non-trivial. A naive approach would be to label the sensor unit as stationary whenever the velocity is (for all practical purposes) zero. However, due to the systematic motions of the foot during the stance phase (see Section \ref{section_modeling_errors}), this will typically lead to too few zero-velocity instances. Thus, a better approach is to label the sensor unit as stationary whenever the velocity measurements fall below some given threshold. However, this means that, once again, we have a threshold that needs to be calibrated. The natural solution is to set this threshold in such a way that the navigation performance is optimized. This idea was explored in \cite{Wagstaff2018}, where several zero-velocity detectors with different fixed thresholds were applied, one by one, to a number of ``motion trials'' (sequences of training data). This included both conventional detectors based on foot-mounted inertial measurements (see Section \ref{section_formalization_zv_detection}) and detectors that utilized velocity estimates from high-accuracy reference systems. For each motion trial, ground truth zero-velocity labels were then extracted from the detector-threshold pair that gave the best positioning performance. To ensure that the best performing fixed-threshold detector is approximately optimal over a given motion trial (so that there is no time-varying threshold that will give a substantially better performance), the environment and gait style were fixed within each individual motion trial. The main limitation of data-driven zero-velocity detectors is the need for large amounts of labelled training data, representative of all intended usage conditions. The collection of such data is both costly and time consuming.

\subsection{Other Robust Detectors}
\label{subsection_other_robust_detectors}

\begin{table*}[t]
    \normalsize
	\caption{Public data sets for development and evaluation of robust zupts. \label{table_data_sets}}
	\centering
	\begin{tabular}{l|ccccccccccc}
		\hline
		\hline
		&  & & & & & \multicolumn{3}{|l|}{} \\ [-2.3ex]
		& \# of & Sampl. & Time & Size of & Ground & \multicolumn{3}{|c|}{Variations of} & Stair & ZUPT \\
		\cline{7-9} 
        Publ. & users & rate $[\text{Hz}]$ & dur. $[\text{min}]$ & area $[\text{m}^2]$ & truth & \multicolumn{1}{|c}{gait style} & surface & \multicolumn{1}{c|}{sens. pos.} & climb. & code \\
		&  \multicolumn{2}{l}{} \\ [-2.7ex]
		\hline    
		&  \multicolumn{2}{l}{} \\ [-2.3ex]
		\cite{Angermann2010} & 3 & 100 & 25 & 25 & VICON & Yes & Yes & No & No & No \\
		\cite{Wagstaff2019} & 5 & 200/125 & \hspace{-1.75mm}120 & 25$^\text{a}$ & VICON$^{\text{b}}$ & Yes & No & No & Yes & Yes \\
		\cite{Wahlstrom2019} & 1 & 250 & 35 & \hspace{-1.75mm}190 & Init. pos.$^{\text{c}}$ & Yes & No & No & No & Yes \\
		\cite{Wahlstrom2020} & 1 & 100 & 60 & \hspace{-1.75mm}225 & Init. pos.$^{\text{d}}$ & Yes & No & Yes & No & Yes  \\
		\hline \hline 
		\multicolumn{11}{l}{\footnotesize{$^{\text{a}}$ Although a lot of data was recorded in a large university building, this data only reflected walking or running along narrow hallways.}} \\ [-0.5ex]
		\multicolumn{11}{l}{\footnotesize{$^{\text{b}}$ As opposed to the data set in \cite{Angermann2010}, this data set only provides position ground truth, not orientation ground truth. In addition, VICON data}} \\ [-0.5ex]
		\multicolumn{11}{l}{\footnotesize{was only provided for the training data. In the testing data, manually surveyed locations were used as ground truth.}} \\ [-0.5ex]
		\multicolumn{11}{l}{\footnotesize{$^{\text{c}}$ All data recordings started and ended at the same position. Thus, performance can be evaluated using the corresponding loop-closure error.}} \\ [-0.5ex]
		\multicolumn{11}{l}{\footnotesize{$^{\text{d}}$ No ground truth was provided for the training data. In the testing data, all data recordings started and ended at the same position.}} \\ [-0.5ex]
	\end{tabular}
	\vspace*{-0mm}
\end{table*}

Several robust zero-velocity detectors that do not fit into the three main categories considered in Sections \ref{section_threshold_adjustment}, \ref{section_gait_cycle}, and \ref{section_data_driven_detectors} have been proposed. Some of these combine multiple conventional detectors \cite{Shaolei2018,Chen2012,Jimenez2010}, whereas others perform the detection based on local optimization methods applied to the likelihood ratio \cite{Liu2018}, or to the norm of the accelerometer or gyroscope signals \cite{Cho2019,Wang2017}. Moreover, it should be noted that, while most adaptive foot-mounted inertial navigation systems concentrate on adaptation of the zero-velocity detection threshold (see Section \ref{section_threshold_adjustment}), it is also possible to adapt other parameters. In \cite{Wang2018}, for example, a gait speed classification was performed using jerk (the derivative of acceleration) and angular velocity. The result of this classification was then used to adjust the window length $w_f+w_b+1$ (see Section \ref{section_formalization_zv_detection}) in the zero-velocity detection. Likewise, instead of adapting certain detection parameters, a conceptually similar idea is to include estimated navigation quantities, such as speed or Euler angles, in the computation of the test statistic (likelihood-ratio) \cite{Abdulrahim2012,Suresh2018,Walder2010}. Finally, instead of using gait cycle segmentation as in Section \ref{section_gait_cycle}, gait cycle models can also be incorporated into zero-velocity detectors by clustering the stance phases based on their time length. In this way, the detector utilizes information on the typical time length of the stance and swing phases \cite{Wang2015,Zhao2015,Muhammad2014}.  

\subsection{Public Data Sets for Robust Zero-Velocity Detection}

There are several public data sets that could be used to develop and evaluate robust zero-velocity detectors. An early example is detailed in \cite{Angermann2010}. This data set contains about 30 minutes of foot-mounted inertial and magnetometer data collected from three users. Reference data was recorded using a camera tracking system, and videos were made available of all data recordings. Furthermore, position and orientation estimates from a ZUPT-aided inertial navigation system were provided; however, no code was made available. The data set considered variations in walking surface, gait style, and gait trajectory. To aid time synchronization, two synchronization events, each consisting of a firm stomp on the ground, were provided at the beginning and end of each data recording. The usefulness of the data was illustrated with a comparative evaluation of four zero-velocity detectors \cite{Angermann2010}. 

Recently, a similar data set was made available \cite{Wagstaff2019}. This data set is divided into training and testing data, and was specifically created for the evaluation of robust zero-velocity detectors. The training data -- consisting of walking, running, stair-climbing, and crawling -- was recorded from one user in a motion capture area of about $5\,\text{m}\times 5\,\text{m}$, and includes ground truth position from a camera tracking system. The testing data is divided into two subsets. The first subset was recorded in the hallways of a university building and includes ground truth at manually surveyed locations. This subset was recorded from five users alternating between walking and running. The second subset was recorded from one user walking up and down a staircase and includes ground truth \emph{vertical} position data on a per-flight basis. In addition, each data recording in the second subset started and ended at the same position, and thus, the position accuracy can be evaluated based on the corresponding loop-closure error. In \cite{Wagstaff2019}, this data set was used to compare the performance of six zero-velocity detectors. Code was provided alongside the data.  

Additional relevant data sets are detailed in \cite{Wahlstrom2019} and \cite{Wahlstrom2020}. The first of these considered walking at two gait speeds with data recorded at a comparatively high sampling rate. The latter considered both three gait speeds and three sensor placements. The data sets were originally used to evaluate a Bayesian zero-velocity detector and an adaptive zero-velocity detector using FootSLAM, respectively. In both data sets, all data recordings started and ended at the same position, and the position accuracy was evaluated based on the corresponding loop-closure error. The data sets are summarized in Table \ref{table_data_sets}.

\section{Why Are There So Few Commercial Systems?}

Considering that foot-mounted inertial navigation has been around for almost two decades and that the technology has the potential to enable a variety of new products and services, the intriguing question arises as to why only a few commercial products have been launched. This question becomes even more puzzling considering that several patents related to the technology were filed as early as the 1990s (see, e.g., \cite{Fyfe1997,Hutchings1995}), and that these patents were followed by a vast amount of research, conducted to improve and refine the technology; see Fig. \ref{figure_bibliometrics}. The answer is certainly multifaceted, but looking back in time the following observations may help to clarify the picture. 

At the beginning of the 21st century, the sensors were just too bulky, power hungry, and costly to enable any viable commercial products; nor was the battery or microcontroller technology mature enough. Not until the smartphone industry had taken the sensor technology to new heights, around 2010, did the sensors become small, power efficient, and cheap enough to be considered for commercial products~\cite{Wahlstrom2017}. Furthermore, it was only a few years into the 2010s that microcontrollers with sufficient computational power (floating point units) and low enough power consumption to realize fully integrated foot-mounted navigation systems with commercially viable operation times ($\gg1$ h) became available. For example, STMicroelectronics and Atmel announced the release of their first micro-controllers with Cortex-M4F cores at the end of 2011. Hence, the 
main obstacle is no longer the hardware per se. Current obstacles holding back the commercialization of the technology are related to i) the lack of ease-of-use and robustness (as highlighted in previous sections of this paper) of the technology, preventing it from being used in safety-critical applications, and ii) for consumer positioning applications, the limited added value that a foot-mounted inertial navigation system brings to a smartphone-embedded system relying upon, for example, GPS and WiFi measurements. That being said, the company Robotic Research recently announced the release of the foot-mounted navigation system WarLoc 
for soldier positioning. Moreover, many commercial foot-mounted sensor systems for human and animal gait analyses have recently started to appear on the market, see e.g., the Gait Up and Werkman Black gait monitoring systems for humans and horses, respectively \cite{Lefeber2019}. These systems are basically foot-mounted inertial navigation systems, but where the main purpose is to extract the dynamics of the individual steps and not the accumulated position change. Thus, the robustness requirements are less stringent.     
      
\section{Summary}

As detailed in the present article, the past fifteen years have seen significant advances within foot-mounted inertial navigation. In particular, researchers have agreed upon a standard inference framework whereby zero-velocity information is integrated as pseudo observations in a nonlinear state-space model and where the problem of zero-velocity detection is formalized as a likelihood-ratio test. As a result of these advances, current zero-velocity-aided inertial navigation systems can, under controlled and stable conditions, produce odometry estimates with a very competitive price-performance ratio. However, the lack of robustness under varying and unforeseen gait and environment conditions still remains one of the main impediments to large-scale commercialization.

\section{Guidance for Future Research}

To advance the research field and increase the technology readiness level of foot-mounted inertial navigation, both the sensors and the algorithms, as well as the evaluation methodology, need to advance further.

As discussed in Section \ref{section_error_sources}, the performance of foot-mounted inertial navigation systems is affected by both sensor errors and modeling errors. As the performance of consumer-grade inertial sensor technology advances --- pushed by the demands of the smartphone and car industries --- the model errors are likely to become the dominant source of error in foot-mounted inertial navigation systems. This will require more refined inertial navigation mechanization schemes that include the earth's rotation rate, use higher-order discretization methods, etc. More importantly, it will require the development of more accurate and robust zero-velocity detectors and zero-velocity measurement models. Next, we will discuss a number of related potential research directions.

As demonstrated by the review in Section \ref{section_adaptive_zupts}, a substantial amount of research effort has gone into developing robust and adaptive zero-velocity detectors. At the same time, research on how to reduce the modeling errors by making the zero-velocity measurement model more realistic is virtually non-existent. This means that the proposed robust foot-mounted inertial navigation systems all still suffer from a number of serious shortcomings that are embedded into the zero-velocity model. One such shortcoming is the binary decision made by the zero-velocity detector. By limiting the detector to a binary output, we lose information about the fluctuating uncertainty of the zero-velocity model. One solution is to reinterpret the nonlinear state-space model \eqref{eq_state_space_model} as having a measurement model that is applied at every sampling instance, but where the measurement error covariance matrix $\mathbf{R}_k$ changes depending on the result of the zero-velocity detection (if the sensor unit is considered to be moving, all elements of the covariance matrix are set to $\infty$). This interpretation then opens up opportunities for implementations where the zero-velocity detector produces a continuous output that is used to scale the measurement error covariance matrix.

As the sensors get better, the need for ZUPTs will diminish. Hence, one way to reduce the impact of the zero-velocity modeling errors is to only apply ZUPTs when the system is actually in need of them. Specifically, the detector need not be implemented to detect all sampling instances where the velocity is close to zero. Rather, in order to ensure satisfactory positioning performance, the detector should balance the need to apply intermittent ZUPTs with the need to avoid significant modeling errors caused by an excessive number of ZUPTs; see e.g., \cite{Wahlstrom2019}. Another way to reduce the impact of the zero-velocity modeling errors is to replace the zero-velocity model with a model learned from data. This could, for example, be achieved by learning a joint zero-velocity detector and zero-velocity measurement function using a Gaussian process state-space model~\cite{SvenssonSchon2017}. This would, among other things, enable the use of \emph{non}-zero velocity updates. However, data-driven methods do not obviate the need for models. In particular, data-driven methods often have trouble extrapolating beyond the training space\footnote{In \cite{Wagstaff2019}, it was remarked that ``the LSTM model was occasionally unable to detect zero-velocity events when a wearer’s speed was outside of the training distribution''.}. Therefore, data-driven methods should, where possible, be complemented or restricted based on available models. Within foot-mounted inertial navigation, one such model could be the rotational symmetry of the inertial measurements. That is, if the evaluation scenario permits the sensor unit to be positioned at an arbitrary orientation with respect to the foot, data-driven functions which map the inertial measurements to navigation data should give the same result regardless of the foot-sensor orientation. One way to heuristically incorporate this constraint is to randomly rotate the training data to simulate different sensor orientations; this approach was taken when learning the motion classifier in \cite{Wagstaff2019}. However, another approach could be to integrate the rotational symmetry into a constrained neural network \cite{Hendriks2020}.

Finally, there is a lack of standardization with regard to the evaluation process. In particular, the research community is in great need of agreed-upon minimum sensor requirements, more public code for baseline implementations, and more diverse and comprehensive reference data sets. The lack of standardization hinders fair and reliable comparison of algorithms, frequently leads to the use of sensors with insufficient bandwidth, dynamic range, etc., and forces researchers to devote time to the development of baseline implementations before being able to focus on the relevant research questions. Thus, to ensure that algorithms are evaluated in an efficient manner, researchers need to build public data sets with a large number of users, strict compliance of sensor requirements, and substantial diversity in both environmental conditions and gait characteristics. All in all, the oversight of this area significantly delays progress, and a joint effort within the research community to address these issues would be of great benefit.

\bibliographystyle{IEEEtran}
\bibliography{refs}

\end{document}